# Generation of Localized Lower-Hybrid Current Drive by Temperature Perturbations

S.J. Frank[1], A.H. Reiman[2], N.J. Fisch[2], P.T. Bonoli[1]

1. Massachusetts Institute of Technology: Plasma Science and Fusion Center, Cambridge, Massachusetts 02139 USA
2. Princeton Plasma Physics Laboratory, Princeton, New Jersey 08544 USA

**Abstract**

Despite high demonstrated efficiency, lower-hybrid current drive (LHCD) has not been considered localized enough for neoclassical tearing mode (NTM) stabilization in tokamaks. This assessment must be reconsidered in view of the RF current condensation effect. We show that an island with a central hot spot induces significant localization of LHCD. Furthermore, in steady state tokamaks where a significant amount of current is provided by LHCD, passive stabilization of NTMs may occur automatically, particularly as islands become large, without requiring precise aiming of the wave power.

## I. Introduction

Disruptions are a major concern in for ITER and future tokamak reactors. One way to address disruptions is through mitigation, i.e. minimizing damage caused by disruptions. Yet disruption mitigation will not alone suffice for dealing with disruptions. Every mitigated disrusption in a reactor class device will cause some damage to the first wall, and it is estimated that ITER will need to maintain a disruption rate of less than 1% to keep cumulative damage to the first wall at an acceptable level. In fusion reactors, unplanned shutdowns will severely impact commercial viability, even if mitigation is 100% successful, and disruptions will need to be extremely rare. Every mitigated disruption will also carry with it some level of risk. It will be desirable to avoid disruptions to the extent possible. The JET device has had a 16% rate of unintended disruptions since it was converted to have an ITER-like wall and 95% of the disruptions are preceded by the appearance of large locked islands [1]. A statistical analysis of disruptions in JET found that there is a distinct island width at which islands cause the tokamak to disrupt corresponding to approximately 30% of the minor radius [2]. This suggests that islands are playing a key role in triggering disruptions. Thus, there is then a critical need for a capability to suppress magnetic islands in tokamaks before they can cause disruptions.

The stabilization of magnetic islands by noninductive rf-driven current has long been predicted [3]. Particularly attention has been given to electron cyclotron current drive (ECCD) [4] and significant progress has been made using this approach, both theoretical [5,6] and experimental [7]. The question remains, however, what is the best means of noninductive rf-driven current to stabilize these islands? There are many RF waves that might be employed to generate toroidal current in tokamaks [8]. The waves that have received the most attention have been the lower hybrid wave for lower hybrid current drive (LHCD) [9] and the electron cyclotron wave for ECCD [4]. Both of these waves have been employed to generate substantial noninductive toroidal current for the purpose of operating tokamaks in the steady state. In this regard, LHCD has been shown to be particularly efficient. On the other hand, for the purpose of NTM stabilization, while LHCD has received some experimental attention [10], nearly all of experimental effort to date has been focused on ECCD. ECCD has been preferred since it is thought to be the only current drive method that can operate in 'in a highly localized, robustly controllable way' [11].

However, the recently identified RF current condensation effect causes RF wave power deposition, that is initially broad, to condense near the center of a magnetic island at high power [12]. This effect relies upon positive feedback; the damping of the power in a magnetic island raises the temperature at the island center relative to the temperature of its edge. For waves with damping rates sensitive to the temperature the wave damping is then increased at the island center. Increased damping at the island center further raises the central island temperature relative to the periphery. The current drive profile then follows the power deposition profile. This effect is pronounced for both LHCD and ECCD, as both of these RF heating and current drive methods have damping which is extremely sensitive to the electron temperature [13, 14]. However, since LHCD normally has both a broader profile and greater temperature sensitivity than ECCD, the current condensation effect should provide relatively greater benefits. Thus, the assessment that LHCD might be too broad for NTM stabilization must be reconsidered in view of the RF current condensation effect.

To perform this assessment, we evaluate the sensitivity of LH power deposition to a temperature perturbation, and the extent to which an elevated central island temperature can localize the damping of lower hybrid waves. The evaluation is carried out by considering lower hybrid waves launched into a model ITER equilibrium, with an assumed temperature perturbation near a rational magnetic surface where islands might form. The ITER equilibrium was chosen as a canonical example of a reactor relevant equilibrium susceptible to 2-1 and 3-2 magnetic islands. What we show is that, for lower hybrid waves launched from the high-field side of the tokamak ("inside launch"), there can be substantial current drive localization for temperature perturbations as large at 15% in even moderately large islands.

Although we report on only a partial scan of all possible parameters, what can already be deduced from the scenarios offered is that, while the LHCD profile remains broader than scenarios offered by ECCD, the localization by the temperature perturbation is clear and pronounced. What this suggests is that a relatively broad profile of LHCD, which might be employed for supplying a significant part of the current in steady state reactors, could act as a passive methodology for controlling the NTM. Passive stabilization stands in contrast to needing to accurately determine the location of the island and then to direct the RF power as needed for ECCD stabilization. In other words, in the presence of broad deposition of LHCD, an emerging island will automatically develop a hot center and condense the lower hybrid driven current so as to passively stabilize the island. In the case of passive stabilization, much of the RF power is not used for the stabilization but might be used for maintaining a steady state. However, a higher degree of localization may be possible when a launcher specifically designed to maximize the effect is used.

The paper is organized as follows: In Sec. II, we discuss properties of the lower hybrid wave and how it triggers the RF condensation effect. In Sec. III, we show with raytracing/Fokker-Planck simulations that the damping in the presence of a temperature perturbation can lead to significantly enhanced power deposition near the local temperature maximum. In Sec. IV, we evaluate the importance of non-Maxwellian effects on the localization of LHCD and RF condensation. In Sec. V, we summarize our main conclusions.

## II. Lower Hybrid Waves

Lower-hybrid current drive (LHCD) has long been employed as an efficient means of current drive and non-inductive sustainment of tokamak discharges. Lower-hybrid waves (LH waves) have frequencies corresponding to the lower-hybrid limit, $\Omega_i \ll \omega \ll \Omega_e$. Here ω is the angular frequency of the lower-hybrid waves and $\Omega_i$ and $\Omega_e$ are the ion and electron cyclotron gyro-frequencies respectively. The lower-hybrid limit typically corresponds to 1-10 GHz frequency window over a wide range of tokamak parameters. LH waves are launched from a waveguide with a slow-wave launching structure which is placed close to the plasma edge in order to ensure good coupling as the lower-hybrid wave is evanescent unless its frequency is below the electron plasma frequency, ω_pe [15]. The waves then propagate until they encounter one of two limits. The first limit, corresponding to LH slow-wave accessibility, is [16,17]:

$$n_{\|} \geq n_{\|acc} = \frac{\omega_{pe}}{\Omega_e} + \sqrt{1 + \left(\frac{\omega_{pe}}{\Omega_e}\right)^2 - \left(\frac{\omega_{pi}}{\omega}\right)^2} \quad (1)$$

Where $n_{\|}$ is the parallel refractive index corresponding to $ck_{\|}/\omega$ and $\omega_{pi}$ is the ion plasma frequency. When the parallel refractive index drops below this limit the wave is reflected and mode converted into a fast wave. The other limit on LH wave propagation is the onset of Landau damping when [16-18]:

$$n_{\|} \geq \frac{5.4}{[T_e(keV)]^{1/2}} \quad (2)$$

When this relation is satisfied the wave is quickly absorbed by non-thermal electron Landau damping at 3-6 times the electron thermal velocity, v_the [18] and drives a plasma current there [4,9]. The non-thermal nature of LH wave damping causes significant distortion of the electron distribution function at high electron energies necessitating a Fokker-Planck calculation to determine the non-linear evolution of the distribution function in response to LHCD and predict the LH wave's absorption and the current drive profiles.

In previous studies of reactor relevant parameter regimes simulations of LHCD have indicated that the current drive profiles should be broad and off axis, between r/a ~0.6-0.8 [19-25], in comparison to the localized current drive which can be obtained with electron-cyclotron current drive (ECCD) [11,26,27]. While LHCD has been suggested as a mechanism for NTM suppression in future tokamak designs no mechanism by which the LHCD could be localized to effectively stabilize the NTMs was described [20,21]. The temperature perturbation associated with thermal insulation in a magnetic island [28-34], however, can be large enough to induce significant localization of the LH wave. Moreover, due to the shape of the temperature perturbation present in these islands, wave damping and therefore current drive is localized near the O-point of the island where the temperature is peaked and the current drive is most effective at suppressing the island [3].

LHCD localization can occur because of the non-thermal nature of LH wave damping. A small increase in electron temperature can increase the number of electrons with $v_e = v_{ph,LH}$ (the phase velocity of the LH wave), by many orders of magnitude inducing strong wave damping, a consequence of the nonthermal electron population available for damping $\propto e^{-\left(\frac{v_{ph,LH}}{v_{the}}\right)^2}$. As a result of the localization, and thus increased heating and current drive within the island, further wave localization can occur as a result of RF-condensation, which is, a non-linear feedback effect that occurs as a result of the RF power deposition balancing with the thermal diffusion [12]:

$$\nabla \cdot [n_e \boldsymbol{\kappa} \cdot \nabla T(\boldsymbol{x})] = -[P_{rf}(T(\boldsymbol{x})) + P_{OH}(T(\boldsymbol{x}))] \quad (3)$$

Where T is the temperature, $n_e$ is the electron density, $\kappa$ is the thermal diffusivity, $P_{rf}$ is the RF heating power, $P_{OH}$ is the ohmic heating power, and $\boldsymbol{x}$ is the spatial coordinate. The evolution of Eq. (3) leads to further peaking of the island temperature profiles about the O-point. The temperature peaking in turn increases LH wave damping, $P_{rf}$, at the island's O-point leading to a feedback loop. This is the RF condensation effect which can be used to further localize LHCD at an island's O-point and greatly increase LHCD's efficiency when used to suppress magnetic islands.

**III. Simulations of LHCD Localization**

Simulations of LHCD were performed using the GENRAY raytracing code [35] and the plasma distribution function's response to the LH wave absorption was modeled using the CQL3D Fokker-Planck code [36,37]. GENRAY models the propagation and absorption of LH waves in the Wentzel-Kramers-Brillouin approximation using raytracing/geometric optics and passes the resulting ray paths to the CQL3D Fokker-Planck code. CQL3D reconstructs the quasi-linear diffusion coefficient along the ray paths then quasi-linearly evolves the distribution function in time and recalculates the damping along the rays. After a sufficient number of timesteps in CQL3D the ray absorption and perturbed distribution function reach a steady state that correctly models the ray damping on a perturbed distribution function. GENRAY/CQL3D provide ray data that can be analyzed on a ray-by-ray basis and current drive profiles that can be used later for stability calculations in order to predict the required launched LH power needed to suppress a magnetic island.

We have calculated how imposed temperature perturbations affect LH power deposition. In practice we are interested in local temperature perturbations produced by the presence of magnetic islands, with their associated change in topology and associated boundary conditions. For our purposes here, of establishing the sensitivity of the power deposition to the temperature perturbation and the associated localization of the power deposition, it is sufficient to consider only temperature perturbations. The coupled problem, with the nonlinear feedback between the temperature perturbation and the power deposition described by Eq. (3), has been left to future work. In order to properly calculate the RF power source term in Eq. (3), the magnetic island's geometry will need to be considered rigorously in the raytracing and Fokker-Planck simulations.

In the following simulations the magnetic equilibrium, temperature profiles, and density profiles, of ITER Scenario 2 [38] generated using TRANSP [39] were used to model LHCD localization. Scenario 2 was chosen as a canonical example of a reactor relevant parameter space susceptible to 2-1 and 3-2 magnetic islands. The temperature, density, and safety factor profiles used in the simulations appear in Figure 1. The ITER scenarios, unlike many other reactor relevant scenarios, are highly vetted, and ITER Scenario 2 has q = 2 and q = 1.5 surfaces that are far enough off-axis that they are accessible to LH waves. ITER scenario 2 also has high, reactor-relevant, electron core temperature, $T_{e0}$, ensuring that the LH wave damps in a single pass. The strong single pass damping of the LH waves in this discharge ensures that the raytracing simulations stay far from the regimes where reflections and edge cut-offs can occur causing the WKB approximation to break down. Additionally, in strong damping regimes, the LH wave is localized to so called "resonance cones" that propagate in an organized fashion [40] as opposed to weakly damped regimes where the wave exhibits a cavity mode like propagation pattern filling the tokamak stochastically [41, 42]. An example of the propagation of the LH waves in Scenario 2 is shown in Figure 2. As LH wave propagation is predictable in strong damping, one can more easily extrapolate the localization associated with a one-dimensional temperature perturbation to the localization that would occur in a more realistic three-dimensional island geometry. This allows one to make an accurate assessment of the LHCD localization expected in a more complex geometry as the wave absorption in the one-dimensional case will be similar to the three-dimensional case assuming the waves hits the island relatively close to the island midplane. The magnetic field perturbation $\delta B$ from the magnetic island is not included in these simulations, but it is expected to have little effect on LH wave propagation as $\delta B$ is much smaller than the background magnetic field $B_0$ [43].

The one-dimensional radial temperature perturbation used in these simulations has the form:

$$T = T_0 \left[1 + \delta T \left(\frac{\rho}{w/2}\right)^2 + \delta T \sin\left(\frac{2\pi\rho}{w/2}\right)\right], \quad -\frac{w}{2} \leq \rho \leq 0$$

$$T = T_0 \left[1 + \delta T \left(\frac{\rho}{w/2}\right)^2\right], \quad 0 < \rho \leq \frac{w}{2} \qquad (4)$$

$$T = T_0, \ otherwise$$

where *w* corresponds to the island width, $\rho$ corresponds to a relative radial coordinate in the magnetic island extending from -w / 2 to w / 2, and $\delta T$ corresponds to a free parameter that allows us to set the perturbation size. The temperature perturbation was centered about either the q = 2 or q = 1.5 flux surfaces. The perturbation in Figure 1, resulting from Eq. (4), is qualitatively similar to the radial temperature perturbations induced by magnetic islands without the presence of local RF heating in previous tokamak experiments, with maximum $\delta T$ values of ~0.05-0.15, as measured using electron-cyclotron emission diagnostics [30-32,34,46]. In future studies where realistic island geometry is added the temperature perturbation would be directly calculated from Eq. (3). The width of the island, w, was set to either 10 cm or 20 cm and the $\delta T$ value associated with the perturbation was varied between 0.05 and 0.15 in steps of 0.05.

In order to ensure LH wave accessibility to the 3/2 and 2/1 rational surfaces it was necessary to launch the LH waves from the high-field side of the Scenario 2 discharge. This is not a realistic launcher configuration for ITER, however, it is relevant to other future reactor design studies with similar parameters such as pulsed or hybrid scenarios in ARC class devices, EU-DEMO, and CFETR [23,24,45]. The high-field side launch was required because the density and temperature pedestal in Scenario 2 causes the LH wave accessibility window defined by Eq. (1) and Eq. (2) to be very small. By increasing the magnetic field at the launch location, and therefore $\Omega_e$, the accessibility constraint imposed by Eq. (1) is relaxed allowing wave accessibility at both the q = 2 and q = 1.5 surfaces. Lower-hybrid waves in the simulations of

Scenario 2 were launched at a frequency of 5GHz from a 0.5 m high waveguide grill positioned 55 degrees above the high-field side midplane with a total launched power of 20 MW. The shape of the LH power spectrum was chosen to be $(\sin(x)/x)^2$ with a peak $n_{||}$ = -1.57 and spectral width $\Delta n_{||}$ = 0.06. The launch $n_{||}$ was chosen after scanning the accessible range of launch $n_{||}$ values because it provided good accessibility to both the q = 2 and q = 1.5 flux surfaces, was capable of inducing localization on both surfaces, and had relatively high phase-velocity which amplifies the localization in response to temperature perturbations. The amplification referred to here happens as a result of damping occurring at higher electron energies. At high energies a smaller $\delta T$ value is required to sufficiently modify the population of superthermal electrons to increase the Landau damping rate and induce localization. Different launcher configurations were found to achieve stronger localization in response to perturbations at one of the two surfaces but were not found to be able to localize effectively at both. In reactor applications multiple launchers with different spectra could be employed for optimized localization on all flux surfaces vulnerable to instabilities.

Results of the coupled GENRAY/CQL3D simulations of Scenario 2 with 20 cm perturbations on the q = 2 surface are shown in Figure 3, and simulations of a 10 cm perturbation on the q = 1.5 flux surface are shown in Figure 4. Localization about the q=2 and q = 1.5 flux surfaces was obtained in the presence of perturbations at all $\delta T$ values. With $\delta T$ values of 0.10 or larger strong localization about the center of the perturbation was obtained. The localization effect was consistent across perturbation widths with localization of power deposition occurring in perturbations with widths of both 10 cm and 20 cm. Substantially better localization, where a majority of the total RF power was deposited within the perturbation half-width, could be achieved when the launcher was optimized for a particular flux surface. An example of such an optimization for localization about the q = 2 flux surface can be seen in Figure 5. As ITER has been predicted by ASTRA simulations of island stabilization with ECCD to have islands with central $\delta T$ values in excess of 0.25 at widths > 20 cm [46], these results have promising implications for reactor relevant stability control with LHCD as they indicate that LHCD is indeed sensitive to temperature perturbations, and could be localized to a magnetic island based only on the temperature perturbation in the island. Localization could allow one to stabilize an island well before it induces a disruption.

Finally, the dependence on location of the LH wave damping relative to the location of the temperature perturbation was examined. In order to achieve a high degree of localization without very large temperature perturbations there must already be some quasi-linear damping of the LH wave at the location where the perturbation is present. If a perturbation is introduced at a location where there is little or no prior LH wave damping, the LH wave damping that results from introducing the perturbation will not be significant unless the perturbation is unrealistically large. For example, in the Scenario 2 simulations when the LHCD launcher was moved to the high-field side midplane the $n_{||}$ evolution experienced by the LH waves while propagating in the toroidal magnetic equilibrium was modified, and as a result the temperature at which the wave damped was increased [41]. After this modification LHCD would no longer localize at the q = 2 flux surface even when a temperature perturbation with $\delta T$ exceeding 0.2 was imposed (however, localization at the 3-2 flux surface was improved). This condition on localization could reduce the viability of LHCD stabilization schemes in steady state scenarios where the locations of the rational surfaces on which islands form do not necessarily correspond to the locations where the current drive is desired. However, if enhancing the effectiveness of RF stabilization with LHCD is considered in the scenario design phase then it is likely that this problem could be overcome. In some cases, this condition is already satisfied too, for example, in ARIES AT [20] a 5-2 NTM at the q = 2.5 flux surface was of some concern as it was unclear whether or not the LHCD there would stabilize it. Based on the LHCD profiles presented in the ARIES AT design, it is likely localization by the temperature perturbation in the island and further localization by RF condensation would occur making LHCD effective at stabilizing NTMs without the need for additional actuators.

## IV. The Importance of Non-Maxwellian Effects in RF Condensation

In the formulation of RF condensation in [1], the $P_{rf}$ term was dependent on $T(x)$. In reality this dependence is tied to the exact details of the electron distribution function i.e. $P_{rf}(f_e(v,x))$ as the slope of the distribution function, $\partial f_e/\partial v$, can profoundly affect the deposition profile of LHCD. To determine if non-Maxwellian, or quasi-linear, damping was indeed important and should be included in future calculations of RF condensation, the simulations of Scenario 2 were examined. Since the electron distribution function becomes more distorted as the absorbed RF power density increases, if the RF power deposition behavior on an initial island temperature perturbation prior to RF condensation is found to be non-Maxwellian, then the higher power densities expected after localization by RF condensation should also exhibit quasi-linear behavior. To examine the quasi-linear dependence of LH wave damping the simulation data was examined ray by ray. The damping rate on the Maxwellian has been compared to the damping rate of that ray on the electron distribution function after it had been evolved by the Fokker-Planck equation. The results of this analysis show modification of the damping rate of rays passing through the perturbation as a result of the formation of a Landau plateau. An example of this for a ray with $n_{\|} = -1.58025$ can be seen in Figure 6.

The formation of a Landau plateau reduces the rays' damping at outer flux surfaces and causes them to penetrate farther into the plasma. This may be a favorable effect since it can prevent "shadowing" of a magnetic island in some cases. Shadowing occurs when the temperature perturbation in a magnetic island as the result of the RF condensation effect becomes large enough that the wave damps before it is able to reach the center of the island. Non-Maxwellian damping should increase the magnitude of the perturbation required for shadowing to occur. Additionally, it was found that when quasi-linear effects were taken into consideration some of the lowest $n_\parallel$ rays demonstrated little response to the temperature perturbations and deposited most of their power further into the plasma. This suggests that it may be possible, with sufficiently broad launched spectrums, to stabilize the 2-1 and 3-2 mode simultaneously. If more accurate island geometry was used in these simulations power densities would be higher as, rather than being spread over the entire volume of the q = 2 flux surface as in the CQL3D simulation, the power would be localized to a much smaller volume in the magnetic island. The use of a 20 MW launch power in the Scenario 2 simulations offsets this inaccuracy somewhat as stabilization of magnetic islands in a pulsed or hybrid scenario would likely not require 20 MW of RF power. The total launched power density in these use cases would be significantly, perhaps even an order-of-magnitude lower, based on simulations of the stabilization of islands in ITER scenario 2 using ECCD [46-48], however, even at lower power densities LHCD exhibits quasi-linear behavior. The inclusion of quasi-linear damping could significantly modify the hysteresis effect LH waves experience as they undergo RF condensation [49]. Though, the modification as a result of quasi-linear effects will likely serve to reduce the amount of edge deposition in the magnetic island and improve the relative increase of current at the island O-point as quasi-linear effects tend to increase the characteristic deposition width of the lower-hybrid wave.

Since we have shown here the power deposition profile can be significantly modified by non-Maxwellian damping, the evolution of the electron distribution function in response to the RF will need to be included in future simulations of LH waves undergoing RF condensation. Therefore, rather than solving a simplified thermal diffusion equation the full Fokker-Planck equation describing the electron distribution function inside of the island will need to be solved due to the strong quasi-linear dependence of the $P_{rf}$ term.

## V. Conclusion

LHCD localization in response to temperature perturbations has been demonstrated in simulations of reactor relevant conditions. For magnetic island-like radial temperature perturbations located at the q = 2 surface with a $\delta T$ values of 0.10 - 0.15, which can reasonably occur in magnetic islands resultant from an NTM before RF condensation feedback, strong localization of LH wave absorption occurred for reactor parameters characteristic of ITER Scenario 2. In all simulations of LHCD localization the majority of the LHCD within the perturbation was located within the half-width, and RF power densities in the perturbed region were more than doubled when the perturbation exceeded $\delta T \sim 0.1$. The strong localization observed in these simulations could be sufficient, in some situations, to stabilize a magnetic island without further localization by RF condensation. Smaller temperature perturbations with $\delta T$ values of 0.05 exhibited modest localization which could reduce the launched power density required to achieve island stabilization. In these simulations the same launch spectrum was used to localize current drive on both rational surfaces for simplicity. Greater localization can be obtained if the launcher configuration is tailored to induce localization on a specific flux surface or multiple launchers with different configurations are used. Finally, if the LH wave were to induce RF condensation in these cases the localization, and therefore island stabilization efficiency, of the LHCD could be increased further.

The non-linear evolution of the electron distribution function in these simulations showed that the effect of non-Maxwellian damping, not considered in previous RF condensation models [1,49], must be included for LH waves. Therefore, future simulations of RF condensation of LH waves should calculate the evolution of the electron distribution function in response to the RF to accurately determine the RF power deposition profiles. While these results have been obtained using 1-D radial temperature perturbations, due to the predictable nature of LH wave propagation in the single pass damping regimes present in these simulations, the results can be extrapolated to a more complicated island geometry and will be used in to inform more sophisticated RF condensation models that calculate the non-linear evolution of the LHCD profiles and temperature within a magnetic island. Iteration of raytracing/Fokker-Planck simulations that precisely calculate damping in the island geometry with simulations of the temperature diffusion in the magnetic island should allow one to demonstrate non-Maxwellian RF condensation of LH waves.

It is worthwhile also to note that the advantageous scenarios were achieved using inside launch lower hybrid waves, i.e. from the high field side of the tokamak. The use of inside launch waves has been contemplated for driving toroidal current in tokamaks such as ARC, and CFETR [23, 45]. However, the use of inside-launch waves in a tokamak reactor, particularly when they drive current away from the tokamak center where magnetic islands tend to reside, carries also the possibility of an advantageous, though speculative, synergistic effect [50,51] of drawing some of the lower hybrid power from the alpha particles through an alpha channeling effect [52].

Even without the further upside associated with inside launch, these results showing strongly localized power deposition already serve to dispel the notion that LHCD is inherently limited to use as broad steady-state current drive actuator and could, with proper scenario and RF system design, be used as localized, high-efficiency, current drive actuator for stability control. Moreover, with careful system design, it is likely that the localization of LHCD and stabilization of NTMs could be achieved nearly passively, i.e. without the need for significant active feedback control. This could have significant implications for the role of LHCD in future fusion experiments and reactors such as CFETR and ARC.

**Acknowledgement**:

This work was supported in part by Scientific Discovery through Advanced Computing Grant Nos. DE-SC0018090 and US DOE Grant Nos.: DE-FG02-8208;91ER54109, DE-AC02-09CH11466, and DE-SC0016072. Thanks to F. Poli for graciously providing us with the ITER scenario 2 profiles used for this study.

**FIGURES:**

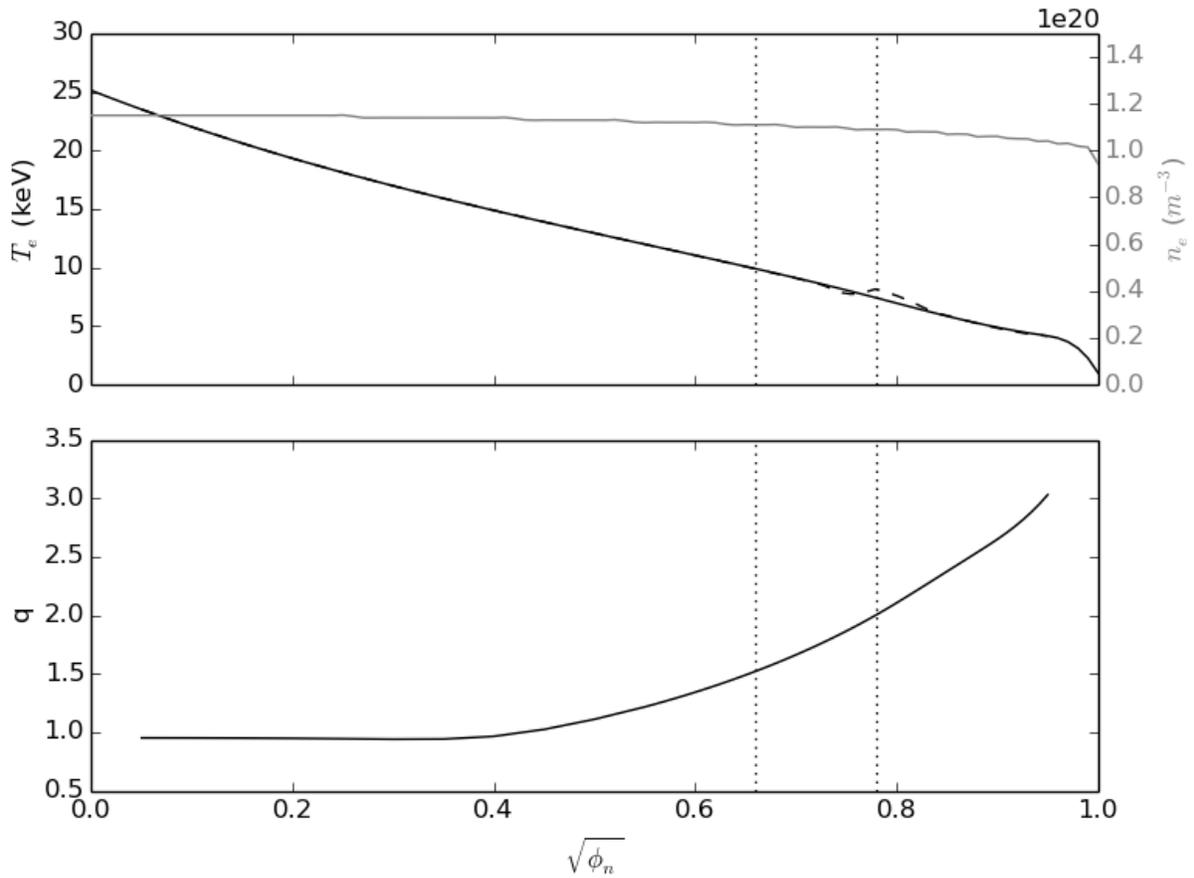

**Figure 1:** ITER Scenario 2 temperature ($T_e$), electron density ($n_e$), and safety factor profiles (q) vs the square root of the normalized toroidal magnetic flux ($\phi_n$). The profiles and magnetic equilibrium were generated using TRANSP [39] and the electron temperatures were then perturbed with a perturbation of the form in Eq. (4) about the q = 2 or q = 1.5 flux surfaces. An example of a perturbation at the q = 2 surface with w = 20cm and $\delta T = 0.10$ is shown by the dashed line in the upper plot.

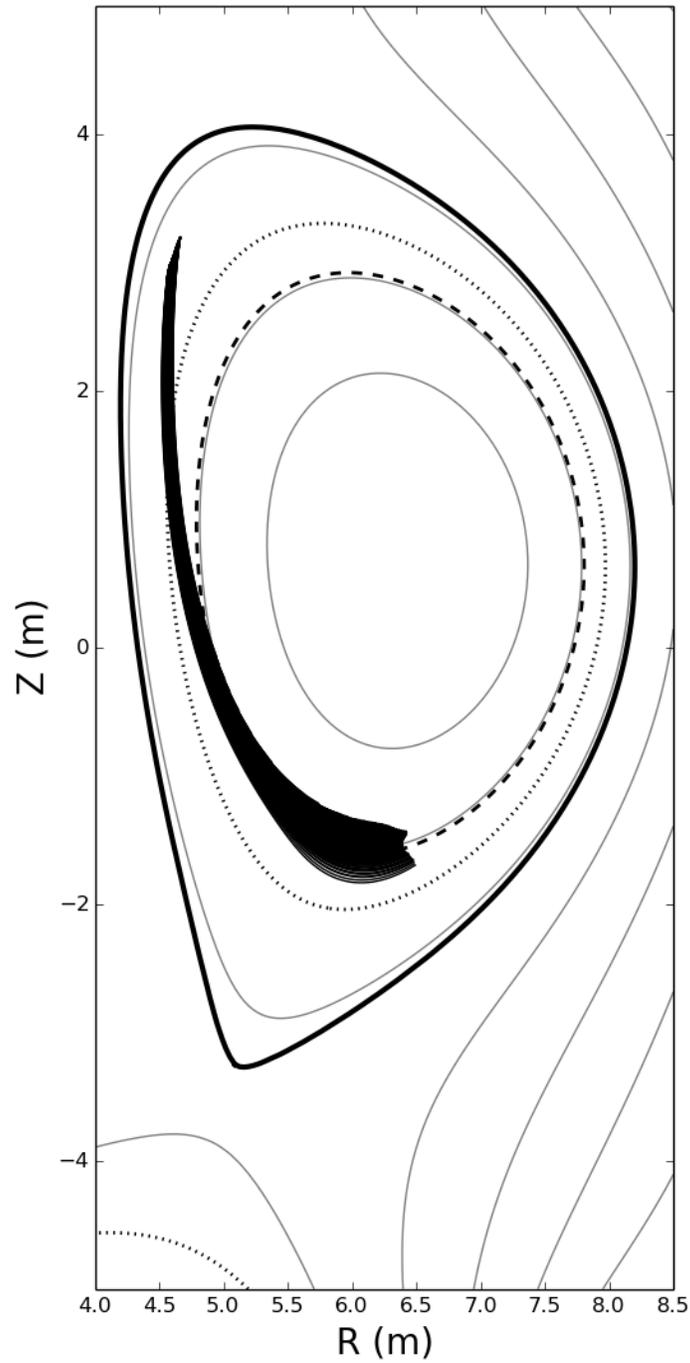

**Figure 2:** Two-dimensional projection of 240 ray paths in an ITER Scenario 2 simulation with no temperature perturbation. The rays shown as bold black lines, propagate in an orderly, predictable, fashion characteristic of strong damping. Rays are launched slightly inside the last closed flux surface at r/a = 0.98. The dotted line indicates the q = 2 surface and the dashed line indicates the q = 1.5 flux surface.

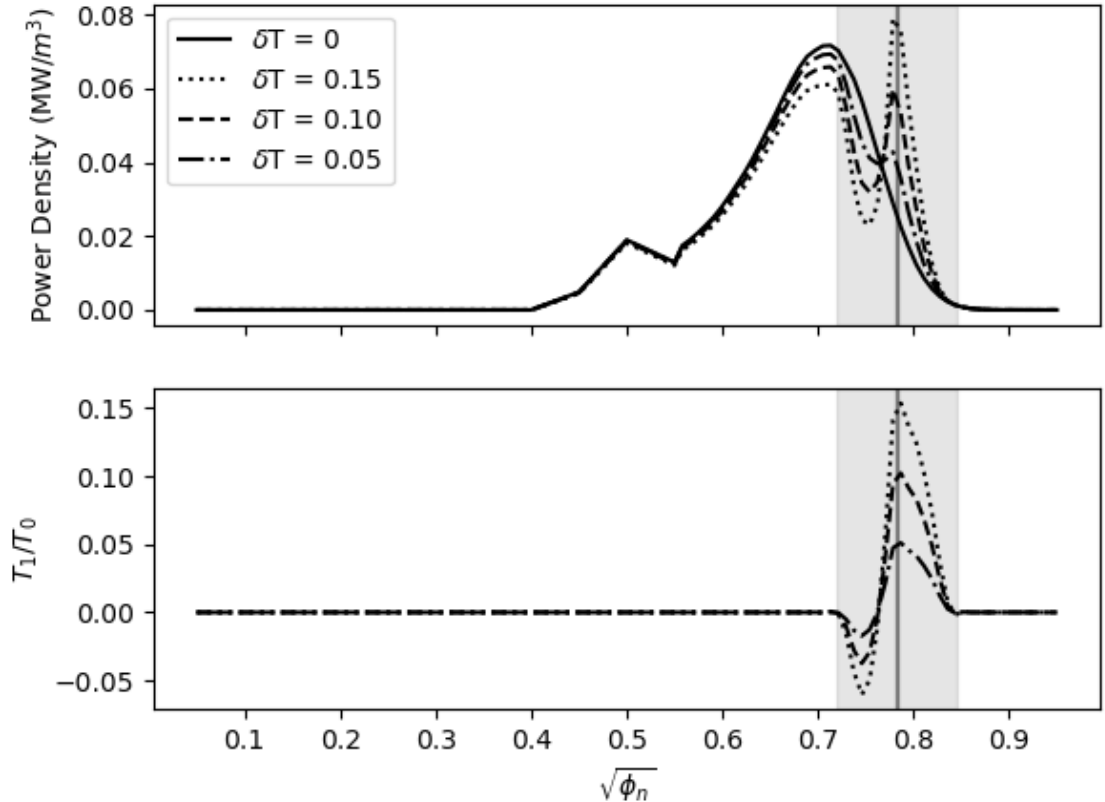

**Figure 3:** Results of a GENRAY/CQL3D simulation showing the power deposition by the lower hybrid wave and $T_1/T_0$ vs. the square root of the normalized toroidal magnetic flux for a 20 cm temperature perturbation centered at the q = 2 flux surface. The vertical line represents the location of the q = 2 flux surface in the simulation and the shaded region represents the region subject to the perturbation. In the case of $\delta T = 0.10$ the power deposited inside the perturbation half width increased from approximately 1 MW to 2 MW and the power density at the center of the perturbation increased by a factor of 2.41.

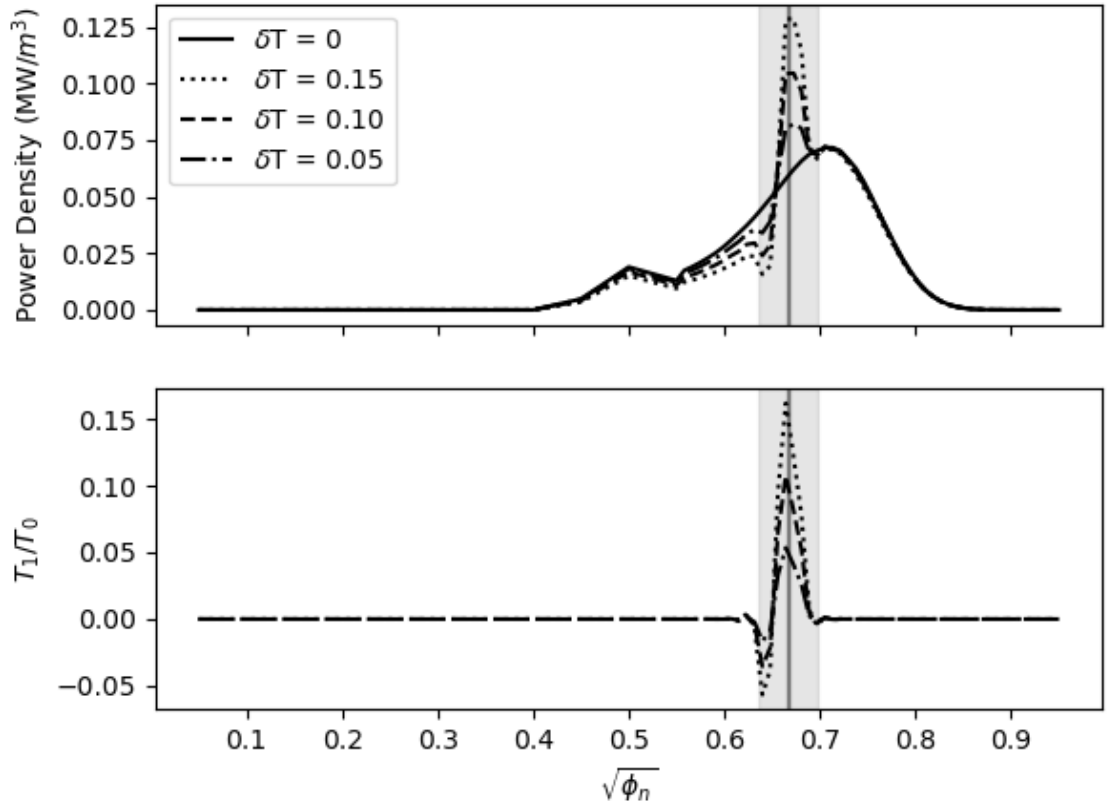

**Figure 4:** Results of a GENRAY/CQL3D simulation showing the power deposition by the lower hybrid wave and $T_1/T_0$ vs. the square root of the normalized toroidal magnetic flux for a 10 cm temperature perturbation centered at the q = 1.5 flux surface. The vertical line represents the location of the q = 1.5 flux surface in the simulation and the shaded region represents the region subject to the perturbation.

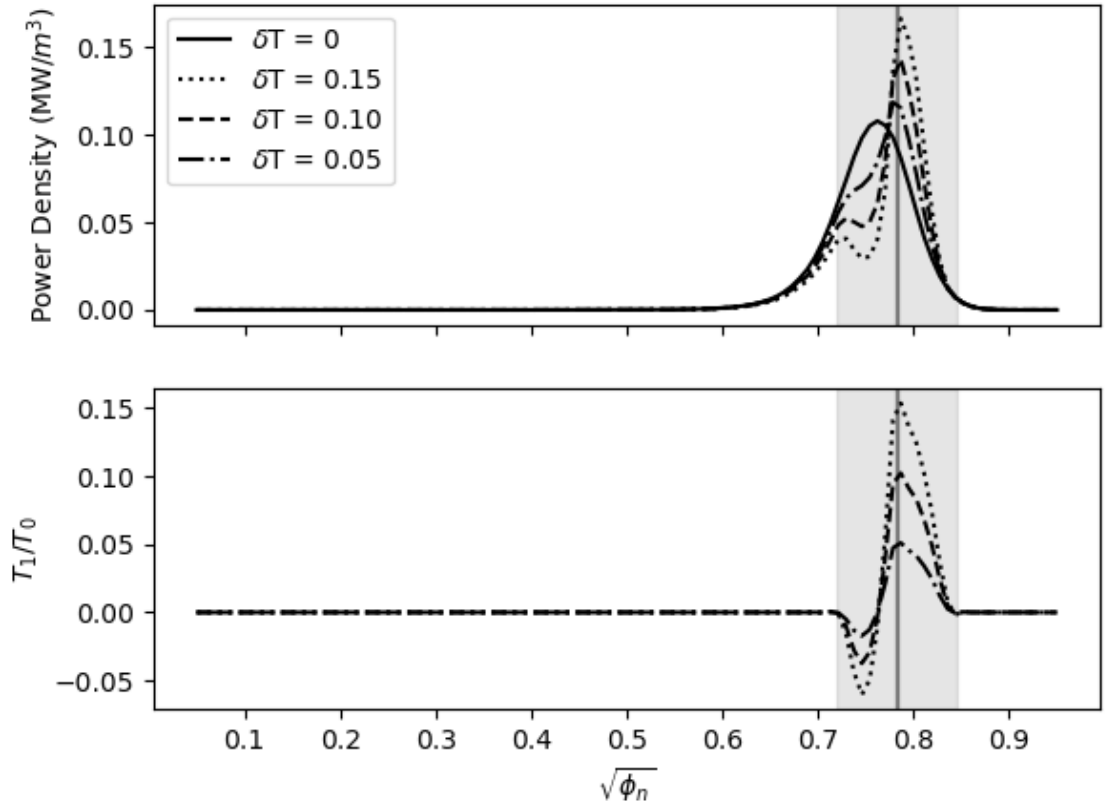

**Figure 5:** Power deposition with a LHCD launcher configuration optimized for q = 2 localization. The LHCD launcher was moved to 60 degrees above the inboard mid-plane and the spectrum centered at $n_\parallel$ = 1.58. All other simulation parameters are the same as those used in Figure 3. After these modifications a majority of the LHCD power is deposited within less than half-width of the perturbation when δT = 0.15.

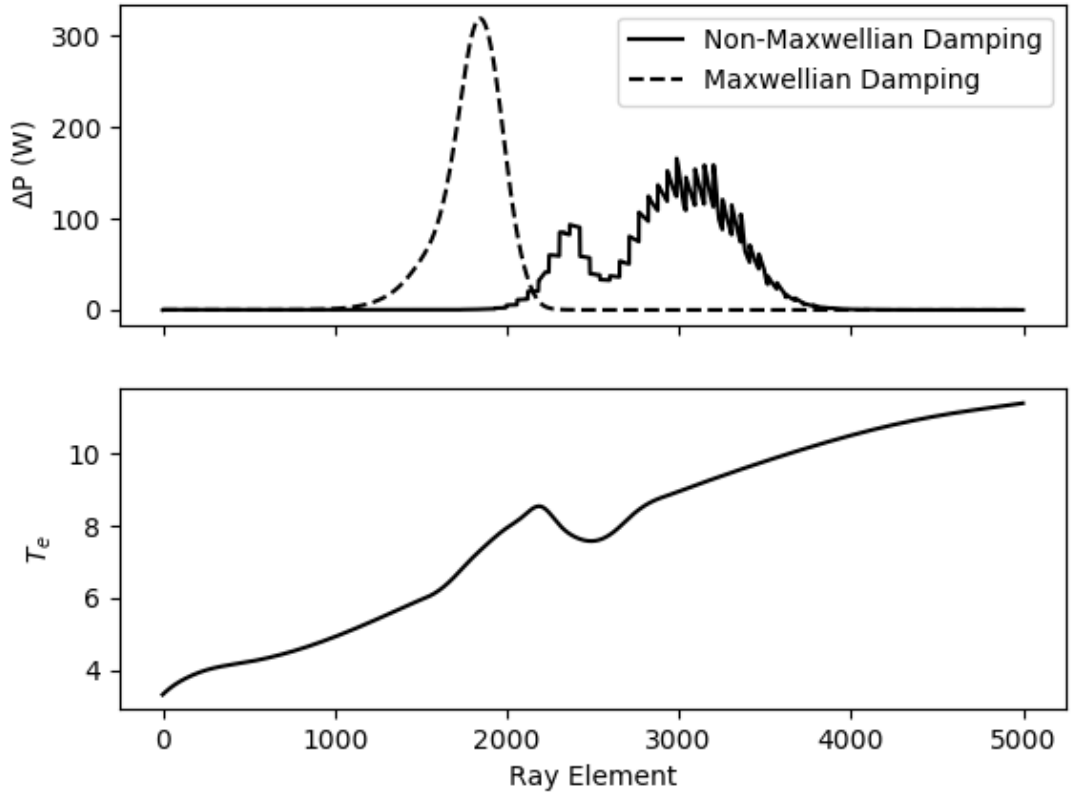

**Figure 6:** Data from a single ray launched with $n_{||}$ = -1.58025 in a GENRAY/CQL3D simulation launched into a perturbed temperature profile with $\delta T$ = 0.15. Plotted for Maxwellian and non-Maxwellian wave absorption is the incremental power deposited at each step along the ray $\Delta P$ and the temperature T along the ray path vs the distance along the ray. The highest phase velocity rays demonstrated very little quasi-linear response to the perturbation. This suggests that both 3-2 and 2-1 modes islands could experience localization simultaneously with a broad launch spectrum.